\documentstyle[12pt]{article}

\newcommand{\be}[1]{\begin{equation}\label{#1}}                     
\newcommand{\ba}[1]{\begin{eqnarray}\label{#1}}                     
\newcommand{\ee}{\end{equation}}                                    
\newcommand{\ea}{\end{eqnarray}}                                    
\newcommand{\non}{\nonumber\\\rule{0pt}{30pt}}
\newcommand{\nona}[1]{\nonumber\\\rule{0pt}{#1pt}}

\newcommand{\dis}{\displaystyle}                                    
\newcommand{\Eq}[1]{(\ref{#1})}                                    
\renewcommand{\Im}{\mathop{\rm Im}} 

\newcommand{\hY}{\hat Y}
\newcommand{\hD}{\hat D}
\newcommand{\co}{{\cal O}}
\newcommand{\fmin}[1]{F_{\mbox{\scriptsize min}}\left(#1\right)}
\newcommand{\ketd}{|0)}
\newcommand{\brad}{(0|}
\makeatletter
\@addtoreset{equation}{section}
\makeatother

\newcommand{\single}{\hphantom{o}}
\begin{document}
%%%%%%%%%%%%%%%%%%%%%%%%%%%%%%%%%%%%%%%%%%%%%%%%%%%%%%%%%%%%%%%%
%
%%%%%%%%%%%%%%%%%%%%%%%%%%%%%%%%%%%%%%%%%%%%%%%%%%%%%%%%%%%%%%%%
%%%%%%%%%%%%%      DOUBLE SPACING BLOCK       %%%%%%%%%%%%%%%%%%
\single
%\double
%%%%%%%%%%%%%%%%%%%%%%%%%%%%%%%%%%%%%%%%%%%%%%%%%%%%%%%%%%%%%%%%%
\begin{flushright}
January 1997 \\
\end{flushright}

\begin{center}
{\large\bf The  determinant representation for
  quantum correlation functions of the sinh-Gordon model}

\vspace{10pt}

{\normalsize 
\bf
V.~E.~Korepin\raisebox{2mm}{$\dagger$} and 
N.~A.~ Slavnov\raisebox{2mm}{$\ddagger$}}\\
\vskip2cm
~\raisebox{2mm}{{\scriptsize $\dagger$}}
{\it ITP, SUNY at Stony Brook, NY 11794-3840, USA.}\\
%e-mail: 
korepin@insti.physics.sunysb.edu\\
%Phone: 1-516-632-7981,~~~FAX: 1-516-632-7954

\vspace{20pt}
~\raisebox{2mm}{{\scriptsize $\ddagger$}}
{\it Steklov Mathematical Institute,
Gubkina 8, Moscow 117966, Russia.}\\
%e-mail: 
nslavnov@mi.ras.ru
%Phone: 7-095-135-1370,~~~FAX: 7-095-135-0555

\end{center}

\vskip24pt

\begin{abstract}
\noindent
We consider the quantum  sinh-Gordon model in this paper. Form 
factors in this model were calculated in \cite{KM}--\cite{ADM}. We 
sum up all contributions of form factors  and obtain a closed 
expression for a correlation function. This expression is a 
determinant of an integral operator. Similar determinant 
representations were proven to be useful not only in the theory of
correlation functions \cite{BMW}--\cite{LLSS}, but also in the matrix models
\cite{D}--\cite{FO}.
\end{abstract}
\vskip4pt
%\begin{center}
%PACS codes: 11.10z, 03.70k, 11.55Ds.\\ 
%Keywords: field theory,exact S-matrices, form factor, correlation
%functions, Fredholm determinant.
%\end{center}

\newpage

%%%%%%%%%%%%%%%%%%%%%%%%%%%%%%%%%%%%%%%%%%%%%%%%%%%%%%%%%%%
\section{Introduction \label{I}}
%%%%%%%%%%%%%%%%%%%%%%%%%%%%%%%%%%%%%%%%%%%%%%%%%%%%%%%%%%%

The theory of massive, relativistic, integrable models is an important
part of modern quantum field theory \cite{ZZ}--\cite{N}. Scattering matrices
in these models  factorize into a product of two-body 
S-matrices \cite{ZZ}. Form factors can be calculated on the basis of
a bootstrap approach \cite{ZZ}--\cite{N}. 

The purpose of this paper is to
calculate correlation functions. As usual correlation function can be
represented as an infinite series of form factors contributions. 
In this paper we sum up all these contributions and obtain a closed 
expression for correlation functions of local operators  
\Eq{DRcordet}.  The idea of this summation is the following. We 
introduce an auxiliary Fock space and auxiliary Bose fields (we shall 
call them dual fields).  These fields help us to represent the form 
factor decomposition of a correlation function in a form similar to 
``free fermionic'' case. This approach was developed in \cite{K}, \cite{S} and
\cite{K.B.I.}. Finally a correlation function is represented as a vacuum
mean value (in the auxiliary Fock space) of a determinant of an integral
operator \Eq{Lkern}. This representation
was proven to be useful \cite{K.B.I.}, \cite{EFIK}, \cite{ES}. It helps for
asymptotical analysis of quantum correlation functions.
Among other things this approach helped to calculate asymptotic of time
and temperature dependent correlation function in Nonlinear Schr\"odinger
equation \cite{KS1}.

In this paper we consider the sinh-Gordon model. It is the model of one
(real) relativistic Bose field $\phi$ in $2D$. The action is
\be{Iaction}
S=\int_{-\infty}^{\infty}\,d^2x\left[\frac12
(\partial_\mu\phi(x))^2-\frac{m_0^2}{g^2}\cosh g\phi(x)\right].
\ee
It is the simplest example of the affine Toda field theories
\cite{MOP} with $Z_2$ symmetry $\phi\to-\phi$.
The model has only one massive  particle. The two-body scattering matrix
 \cite{AFZ}, \cite{AK} is given by an expression
\be{ISmat}
\begin{array}{l}
{\dis
S(\beta,B)=\frac{\tanh\frac12\left(\beta-\frac{i\pi B}2\right)}
{\tanh\frac12\left(\beta+\frac{i\pi B}2\right)},}\non
{\dis
B=\frac{2g^2}{8\pi+g^2}.}
\end{array}
\ee
We shall consider a real $g$, which corresponds to a positive
$B$. Later we shall use a variable
\be{Ibeta}
x=e^\beta
\ee
instead of rapidity $\beta $ .

We shall use a representation for form factors found  in 
\cite{KM}--\cite{ADM} (another representation can be found in  
\cite{L}) 
\be{Iff} 
F_n(\beta_1,\dots,\beta_n)=\langle0|\co(0)|\beta_1,\dots,\beta_n\rangle
=H_nQ_n(s)\prod_{i>j}^n
\frac{\fmin{\beta_{ij}}}{x_i+x_j}.
\ee
Here $\beta_{ij}=\beta_{i}-\beta_{j}$. A function $\fmin{\beta}$ is
holomorphic for  real $\beta$
\be{Ifmin}
\fmin{\beta}={\cal N}(B)\Xi(\beta),
\ee
where
\be{IXi}
\Xi(\beta)=\exp\left[8\int_0^\infty\frac{dx}x
\frac{\sinh\left(\frac14xB\right)\sinh\left(\frac12x\Bigl(1-\frac12B
\Bigr)\right)\sinh\frac12x}{\sinh^2x}\sin^2\left(
\frac{x\hat\beta}{2\pi}\right)\right],
\ee
\be{IcalN}
{\cal N}(B)=\exp\left[-4\int_0^\infty\frac{dx}x
\frac{\sinh\left(\frac14xB\right)\sinh\left(\frac12x\Bigl(1-\frac12B
\Bigr)\right)\sinh\frac12x}{\sinh^2x}
\right],
\ee
and $\hat\beta=i\pi-\beta$. The function $\fmin{\beta}$ has a simple zero
at $\beta=0$ and  no poles at the strip $0\le\Im\beta\le\pi$.
At $\beta\to\infty$ it goes to one: $\fmin{\beta}\to1$.
 Functions $Q_n(x_1,\dots,x_n)$ are symmetric polynomials of
variables $x_1,\dots,x_n$ given by \cite{KM}
\be{IQn}
Q_n(s)={\det}_{n-1}M_{ij}(s),\qquad i,j=1,\dots,n-1,
\ee
where
\be{Imjk}
M_{ij}(s)=\sigma_{2i-j}[i-j+s].
\ee
Let us explain the notations. Here and later we suppress the dependency
of $Q_n(s)$ on variables $x_j$. The index  $n-1$ in the expression
${\det}_{n-1}$ denotes the dimension of the matrix $M_{ij}(s)$.
The functions $\sigma_k$ are elementary symmetric polynomials of the
$k$-th order of the variables $x_1,\dots,x_n$:
\be{Ielsympol}
\sigma_k\equiv\sigma_k^{(n)}(x_1,\dots,x_n)
=\sum_{i_1<i_2<\dots<i_k}^nx_{i_1}x_{i_2}\dots x_{i_k},
\ee
and $\sigma_k=0$ if $k<0$ or $k>n$. Here we also  suppressed the
dependency of $\sigma_k$ on $x_j$. The symbol $[m]$ is a 
``$q$-number'' defined by
\be{Iqnumb}
[m]=\frac{\sin \frac {mB}2}{\sin\frac B2}=\frac{q^m-q^{-m}}{q-q^{-1}},
\ee
where $q=\exp\{iB/2\}$. The number $s$ in \Eq{IQn}, \Eq{Imjk} is an
arbitrary integer, depending on specific choice of an operator $\co$
in \Eq{Iff}.

Finally, the constants $H_k$ in \Eq{Iff} are normalization constants
\be{Iconst}
H_{2n+1}=H_1\mu^n,\qquad
H_{2n}=H_0\mu^n,\qquad
\mu=\frac{4\sin\frac{\pi B}2}{\fmin{i\pi}},
\ee
where $H_0$ and $H_1$ also  depend on specific
operator $\co$. For instance, the form factor of local field is given
by \Eq{Iff} with $s=0$ and
\be{Inorm}
\begin{array}{l}
{\dis H_0=\langle0|\phi(0)|0\rangle=0,}\non
{\dis H_1=\langle0|\phi(0)|\beta\rangle=\frac1{\sqrt{2}}.}
\end{array}
\ee

A correlation function of an operator $\co$ can be presented as
an infinite series of form factors contributions
\be{Icorff}
\langle0|\co(0,0)\co(x,t)|0\rangle=
\sum_{n=0}^\infty \int\frac{d^n\beta}{n!(2\pi)^n}
|F_n(\beta_1,\dots,\beta_n)|^2\prod_{j=1}^ne^{-mr\cosh\beta_j}.
\ee
In the present paper we sum up this series explicitly. 
Now let us present a  plan of the paper. Section \ref{TFF}
is devoted to a transformation of the determinants \Eq{IQn},
\Eq{Imjk} to a form, which is convenient for summation. In 
section 3 we introduce auxiliary quantum operators---dual fields---in
order to factorize an expression for a correlation function and to represent
it in a form similar to ``free fermionic case". In  section 4 we sum 
up the series \Eq{Icorff} into a Fredholm determinant. In  section 5 
we use the Fredholm determinant representation 
for derivation of an asymptotic behavior of  correlation functions.

%%%%%%%%%%%%%%%%%%%%%%%%%%%%%%%%%%%%%%%%%%%%%%%%%%%%%%%%%%%
\section{A transformation of the form factor \label{TFF}}
%%%%%%%%%%%%%%%%%%%%%%%%%%%%%%%%%%%%%%%%%%%%%%%%%%%%%%%%%%%

A  determinant of a linear integral operator $I+V$ can be written
as
\be{TdefFrdet}
\det(I+V)=\sum_{n=0}^\infty\int\,\frac{dx_1\cdots dx_n}{n!}
{\det}_n\left(
\begin{array}{ccc}
V(x_1,x_1)&\cdots&V(x_1,x_n)\\
V(x_2,x_1)&\cdots&V(x_2,x_n)\\
\cdot&\cdot&\cdot\\
V(x_n,x_1)&\cdots&V(x_n,x_n)
\end{array}\right).
\ee
Thus, in order to obtain a determinant representation for correlation
functions one need to represent the form factor  expansion
\Eq{Icorff} in the form \Eq{TdefFrdet}. Determinants of integral operators,
which we consider also can be called Fredholm determinants.

The form factors \Eq{Iff} are proportional to polynomials $Q_n(s)$, which
in turn are equal to determinants of $(n-1)\times(n-1)$ matrices \Eq{IQn}
\be{TQn}
Q_n(s)={\det}_{n-1}M_{ij}(s).
\ee
The matrix $M_{ij}(s)$ consists of $(n-1)^2$ different functions,
depending on the same set of arguments $x_1,\dots,x_n$:
\be{Tmjk}
M_{ij}(s)=\sigma_{2i-j}[i-j+s],\qquad i,j=1,\dots,n-1.
\ee
The main goal of this and  next section is to transform the matrix
\Eq{Tmjk} to  such a form, that entries of a new matrix will be 
parameterized by a single function, depending on different sets of 
variables, (like $V(x_i,x_j)$ in \Eq{TdefFrdet})

\be{Tidea}
M_{ij}\to \hD_{ij}, \qquad \hD_{ij}=\hD(x_i,x_j).
\ee
Here $\hD(x,y)$ is a  function of two arguments.
The element $\hD_{ij}$ depends on $i$ and  $j$ only by means of its
arguments $x_i$ and $x_j$.

First, it is useful to rewrite the representation \Eq{TQn} in terms
of a determinant of a  matrix $n\times n$  . To do this, notice that
$\sigma_{2n-j}=0$, if $j < n$, so  $M_{nj}=\delta_{nj}
[s]\prod_{m=1}^nx_m$. Thus, we obtain
\be{TQn1}
Q_n(s)=[s]^{-1}\prod_{m=1}^nx_m^{-1}{\det}_{n}M_{ij}(s), 
\qquad i,j=1,\dots,n.
\ee
The r.h.s. of \Eq{TQn1} is well defined for all $s\ne0$ and $n\ne 0$. These
two cases should be considered separately.
It is easy to see that for $s=0$ and $n\ne0$ one have to
 understand \Eq{TQn1} as a
limit $s\to 0$, because the determinant is  proportional to $[s]$.
However, for $n=0$, the original representation
\Eq{TQn} is not well defined, while it is natural to define the determinant
${\det}_0M_{ij}=1$ in  \Eq{TQn1}. So, we obtain $Q_0(s)=[s]^{-1}$
for $s\ne0$. On the other hand the case $s=0$ corresponds to the 
form factor of local field.  In this case  we have $H_0=0$, and  the 
form factor is equal to zero $ H_0 Q_0(0) =F_0=0$. Thus, we define
$Q_0(s)=[s]^{-1}$ for $s\ne0$. We do not define $Q_0(0)$, but
we simply put $F_0=0$ for $s=0$.

In  order to study correlation functions we need to find the
square of polynomials $Q_n(s)$.
\be{Tsquare}
Q_n^2(s)=[s]^{-2}\prod_{m=1}^nx_m^{-2}{\det}_{n}C_{jk},
\ee
where
\be{TC}
C_{jk}=(M^T\cdot M)_{jk}=\sum_{i=1}^n
[i-j+s][i-k+s]\sigma_{2i-j}\sigma_{2i-k}.
\ee
One can calculate the sum in \Eq{TC} using an integral representation
for elementary symmetric polynomials
\be{Tsympol}
\sigma_k=
\frac1{2\pi i}\oint\frac {dz}{z^{n-k+1}}
\prod\limits_{m=1}^n(z+x_m).
\ee
Here the integral is taken in positive direction
 with respect to an arbitrary circle $|z|=\rho$ around the
origin. Choosing the radius of the circle $\rho>1$ and
using \Eq{Eint1}  from Appendix A, we find
\ba{Tintrep}
&&{\dis\hspace{-1cm}
C_{jk}=\frac1{(2\pi i)^2}\oint\,d^2z\frac{\prod\limits_{m=1}^n
(z_1+x_m)(z_2+x_m)}{(q-q^{-1})^2}
z_1^{n-j+1}z_2^{n-k+1}}\non
&&{\dis
\times\left\{
\frac{q^{2n+2+2s-j-k}}{q^2z_1^2z_2^2-1}+
\frac{q^{-2n-2-2s+j+k}}{q^{-2}z_1^2z_2^2-1}-
\frac{q^{k-j}}{z_1^2z_2^2-1}-
\frac{q^{j-k}}{z_1^2z_2^2-1}\right\}.}
\ea
In order to get a common denominator we make  replacements of
variables in the braces:
$z_1q^{1/2}=w_1$, $z_2q^{1/2}=w_2$ in the first term;
$z_1q^{-1/2}=w_1$, $z_2q^{-1/2}=w_2$ in the second term;
$z_1q^{1/2}=w_1$, $z_2q^{-1/2}=w_2$ in the third term and
$z_1q^{-1/2}=w_1$, $z_2q^{1/2}=w_2$ in the fourth term.
After simple algebra we arrive at
\be{TrepG}
C_{jk}=\frac1{(2\pi i)^2}\oint\,d^2w\frac{
w_1^{n-j+1}w_2^{n-k+1}}{w_1^2w_2^2-1}
G^{(j)}(w_1)G^{(k)}(w_2),
\ee
where
\be{TG}
G^{(\ell)}(w)=\frac1{q-q^{-1}}\left(
q^{s+\frac{n-\ell}2}\prod_{m=1}^n(wq^{-1/2}+x_m)-
q^{-s-\frac{n-\ell}2}\prod_{m=1}^n(wq^{1/2}+x_m)\right).
\ee

The  matrix $C$ still depends on $n^2$ different functions
$C_{jk}$. However, this matrix can be  transformed to a more convenient
form. Let us introduce  matrix $A_{jk}$ (it is studied in Appendix B)
\be{TA}
A_{jk}=\frac1{(n-j)!}\left.\frac{d^{n-j}}{dx^{n-j}}
\prod\limits_{m\ne k}^n\left(x+x_m\right)
\right|_{x=0},
\ee
with a determinant
\be{TdetA}
\det A=\prod_{a<b}^n(x_a-x_b).
\ee
 Instead of matrix $C$ it  will be convenient to introduce matrix $D$
\be{TD}
D=A^TCA.
\ee
Determinants of matrices $C$ and $D$ are related by
\be{TCD}
{\det}_nC=\prod_{a>b}^n(x_a-x_b)^{-2}{\det}_nD,
\ee

The calculation of the explicit expression for matrix $D$
in \Eq{TD} reduces to the summation
of the Taylor series (see \Eq{SUTs}), so we have
\be{TDjk}
D_{jk}=\oint\,d^2w\frac{w_1w_2}{(2\pi i)^2(w_1^2w_2^2-1)}
Y(w_1,x_j)Y(w_2,x_k),
\ee
where
\be{TY}
Y(w,x)=\frac{J(w)}{q-q^{-1}}
\left(\frac{q^s}{wq^{1/2}+x}-\frac{q^{-s}}{wq^{-1/2}+x}\right),
\ee
and
\be{TJ}
J(w)=\prod\limits_{m=1}^n(wq^{1/2}+x_m)(wq^{-1/2}+x_m).
\ee
Taking the integral, for instance,  with respect to $w_2$ 
(recall that $|w_1w_2|>1$), we have after the symmetrization of the
integrand
\be{TDjk1}
D_{jk}=\frac1{8\pi i}\oint\frac{dw}{w}
\Bigl(Y(w,x_j)+Y(-w,x_j)\Bigr)\Bigl(Y(w^{-1},x_k)+Y(-w^{-1},x_k)\Bigr).
\ee
Thus, we obtain a new representation for the square of the polynomial
$Q_n(s)$:
\be{TQnrep}
Q_n^2(s)=\frac{{\det}_nD}
{[s]^2\prod\limits_{m=1}^nx_m^2\prod\limits_{a>b}^n(x_a-x_b)^2}.
\ee
This brings us closer to \Eq{TdefFrdet}.
%%%%%%%%%%%%%%%%%%%%%%%%%%%%%%%%%%%%%%%%%%%%%%
\section{Dual fields \label{DF}}
%%%%%%%%%%%%%%%%%%%%%%%%%%%%%%%%%%%%%%%%%%%%%%%

The entries of the matrix $D_{jk}$ are parameterized now by a single function
$D$ \Eq{TDjk1}. However, an element $D_{jk}$,  is still  not a
 function of two arguments only, because of  the product
$J(w)=\prod_{m=1}^n(wq^{1/2}+x_m)(wq^{-1/2}+x_m)$. This product depends on 
all  $x_m$. In order to get rid of these products we
introduce auxiliary Fock space and auxiliary  quantum operators---dual fields.
Dual fields are linear combinations of canonical Bose fields, see page 210
of \cite{K.B.I.} .

Let us define
\be{Ddef}
\begin{array}{l}
\dis \Phi_1(x)=q_1(x)+p_2(x),\nona{23}
\dis \Phi_2(x)=q_2(x)+p_1(x),
\end{array}
\ee
where operators $p_j(x)$ and $q_j(x)$ act in the canonical Bose Fock space 
in a following way
\be{Dact}
\brad q_j(x)=0, \qquad p_j(x)\ketd=0.
\ee
Non-zero commutation relations are given by
\be{Dcom}
{}[p_1(x),q_1(y)]=[p_2(x),q_2(y)]=\xi(x,y)=
\log\Bigl((x+yq^{1/2})(x+yq^{-1/2})\Bigr).
\ee
Due to the symmetry of the function $\xi(x,y)=\xi(y,x) $,
all fields $\Phi_j(x)$ commute with each other
\be{DAbel}
{}[\Phi_j(x),\Phi_k(y)]=0,\qquad j,k=1,2.
\ee
However, despite of these simple commutation relations, the vacuum
expectation value of the  dual fields may be non-trivial,
for example:
\be{Dexam}
\brad\Phi_1(x)\Phi_2(y)\ketd=\brad p_2(x)q_2(y)\ketd=\xi(x,y).
\ee
It is easy to show that an exponent of dual field acts like a shift
operator. Namely, if $f\left({\Phi_1(y)}\right)$ is a
function of $\Phi_1(y)$ then
\ba{Dvac1}
&&{\dis
\brad\prod_{m=1}^ne^{\Phi_2(x_m)}f\left({\Phi_1(y)}\right)\ketd
=\brad\prod_{m=1}^ne^{p_1(x_m)}f\left({q_1(y)}\right)\ketd}\non
&&{\dis
=\brad f\left({q_1(y)+\sum_{m=1}^n\xi(x_m,y)}\right)\ketd
=f(\log J(y)).}
\ea
Using this property of dual fields one can remove the products
$J(w)$ from the matrix $D_{jk}$.

Let us define
\be{DhY}
\hY(w,x)=\frac{e^{\Phi_1(w)}}{q-q^{-1}}
\left(\frac{q^s}{wq^{1/2}+x}-\frac{q^{-s}}{wq^{-1/2}+x}\right),
\ee
and
\be{TDjk2}
\hD_{jk}=\frac1{8\pi i}\oint\frac{dw}{w}
\Bigl(\hY(w,x_j)+\hY(-w,x_j)\Bigr)
\Bigl(\hY(w^{-1},x_k)+\hY(-w^{-1},x_k)\Bigr).
\ee
Then, due to \Eq{Dvac1}, we have
\be{DdetD}
{\det}_nD=\brad\prod_{m=1}^ne^{\Phi_2(x_m)}{\det}_n\hD\ketd,
\ee
or
\be{DdetD1}
{\det}_nD=\brad
{\det}_n\left(\hD(x_j,x_k)e^{\frac12\Phi_2(x_j)+\frac12\Phi_2(x_k)}
\right)\ketd.
\ee
The entries of the matrix $\hD_{jk}$ depend on $x_j$ and
$x_k$ only, and they do not depend on other variables $x_m$. Thus,
we have presented the square of the polynomial $Q_n(s)$ in terms
of a vacuum expectation value of a  determinant of
a  matrix  $n\times n$, similar to one of the terms in the r.h.s of
\Eq{TdefFrdet}.
 Entries of matrix $D$ are parameterized by the single
two-variable function $\hD(x,y)$. Let us emphasize again that 
as  an operator in the auxiliary Fock space  $\hD(x,y)$ belongs to an
Abelian sub algebra.

Besides the polynomial $Q_n(s)$ the form factor \Eq{Iff} is  proportional to
a double product $\prod_{a>b}^n\fmin{\beta_{ab}}(x_a+x_b)^{-1}$. 
In order to transform \Eq{Icorff} to \Eq{TdefFrdet} it is  necessary to
 factorize this product.
To do this we introduce  another dual field
\be{Ddef0}
\tilde\Phi_0(x)=\tilde q_0(x)+\tilde p_0(x).
\ee
As usual
\be{Dact0}
\brad \tilde q_0(x)=0, \qquad \tilde p_0(x)\ketd=0.
\ee
Operators $\tilde q_0(x)$ and $\tilde p_0(y)$  commute with all  $p_j$ and 
$q_j$ ($j=1,2$). The only non-zero  commutation relation is 
\be{Dcom0}
{}[\tilde p_0(x),\tilde q_0(y)]=\eta(x,y),
\ee
where
\be{Deta0}
\eta(x,y)=\eta(y,x)
=2\log\left|\frac{\fmin{\log\frac xy}}{x^2-y^2}\right|.
\ee
Here we have used the fact that $|\fmin{z}|=|\fmin{-z}|$. It
is worth mentioning also that the r.h.s. of \Eq{Deta0} has no
singularity at $x=y$, because $\fmin{x}$ has the first order zero at
$x=0$ and $F'_{\mbox{\scriptsize min}}(0)=
\left(i\sin \frac{\pi B}2\fmin{i\pi}\right)^{-1}$ (see \cite{KM}). Hence
\be{Detaxx}
\eta(x,x)=-2\log\left|2x^2\sin \frac{\pi B}2\fmin{i\pi}\right|.
\ee

Newly introduced dual fields also commute

\be{DAbel0}
{}[\tilde\Phi_0(x),\tilde\Phi_0(y)]=0=
[\tilde\Phi_0(x),\Phi_j(y)].
\ee
However, due to the Campbell--Hausdorff formula, we have
\be{Dvac0}
\brad\prod_{m=1}^ne^{\tilde\Phi_0(x_m)}\ketd=
\prod_{a,b=1}^ne^{\frac12\eta(x_a,x_b)}
=\lambda^{-n}
\prod_{m=1}^nx_m^{-2}\prod_{a>b}^n
\left|\frac{\fmin{\log\frac {x_a}{x_b}}}{x_a^2-x_b^2}\right|^2,
\ee
where
\be{DRxi}
\lambda=|2\sin\frac{\pi B}2\fmin{i\pi}|.
\ee
Combining the last formula and the representations \Eq{TQnrep},
\Eq{DdetD1} for $Q_n^2(s)$ we find
\be{Dfactfmin}
Q_n^2(s)\prod_{a>b}^n
\left|\frac{\fmin{\log\frac {x_a}{x_b}}}{x_a+x_b}\right|^2=
\frac{\lambda^n}{[s]^2}
\brad{\det}_n\hat V(x_j,x_k)\ketd.
\ee
Here
\be{DV}
\hat V(x_j,x_k)=\hD_{jk}e^{\frac12\Phi_0(x_j)+\frac12\Phi_0(x_k)},
\ee
and
\be{DnewPhi}
\Phi_0(x)=\tilde\Phi_0(x)+\Phi_2(x).
\ee
So we managed to represent a square of an  absolute value of the
form factor as a determinant, similar to one of the terms in the r.h.s of
\Eq{TdefFrdet}. In the next section we shall 
sum up all contributions of the  form factors and obtain a determinant 
representation for a correlation function. 
%%%%%%%%%%%%%%%%%%%%%%%%%%%%%%%%%%%%%%%%%%%%%%
\section{The determinant representation for a 
correlation function \label{DRCF}}
%%%%%%%%%%%%%%%%%%%%%%%%%%%%%%%%%%%%%%%%%%%%%%

In the previous sections we have obtained the 
representation for a square of an absolute value of the form factor
\be{DRdefff}
F_n(\beta_1,\dots,\beta_n)=\langle0|\co(0,0)|
\beta_1,\dots,\beta_n\rangle,
\ee
in terms of a determinant
\be{DRdetff}
|F_n(\beta_1,\dots,\beta_n)|^2=\frac1{[s]^2}|H_n|^2\lambda^n
\brad {\det}_n\hat V(x_j,x_k)\ketd,
\ee
Here constants $H_n$ are equal to
\be{DRconst}
H_{2n+1}=H_1\mu^n,\qquad
H_{2n}=H_0\mu^n,\qquad
\mu=\frac{4\sin\frac{\pi B}2}{\fmin{i\pi}},
\ee
and
\be{DRh0h1}
\frac1{[s]}H_0=F_0=\langle0|\co(0,0)|0\rangle,\qquad
H_1=F_1=\langle0|\co(0,0)|\beta\rangle.
\ee

We have the representation for a correlation function of operators $\co$
in terms of form factors:
\be{DRcorff}
\langle0|\co(0,0)\co(x,t)|0\rangle=
\sum_{n=0}^\infty \int\frac{d^n\beta}{n!(2\pi)^n}
|F_n(\beta_1,\dots,\beta_n)|^2\prod_{j=1}^ne^{-\theta(x_j)},
\ee
where
\be{DRtheta}
\theta(x)=\frac{mr}2(x+x^{-1}).
\ee
Substituting here \Eq{DRdetff}--\Eq{DRconst} we arrive at a following
representation
\newpage
\ba{DRcordetn}
&&{\dis\hspace{-1cm}
\langle0|\co(0,0)\co(x,t)|0\rangle
=\brad\frac1{[s]^2}\left\{\frac{|H_0|^2+|H_1|^2|\mu|^{-1}}2
\sum_{n=0}^\infty \int_0^\infty
\frac{d^nx}{n!}\left(\frac{|\lambda\mu|}{2\pi}\right)^n
\right.}\non
&&{\dis\hspace{3.5cm}\times
{\det}_n\left[\frac{\hat V(x_j,x_k)}{\sqrt{x_jx_k}}
e^{-\frac12(\theta(x_j)+\theta(x_k))}\right]}\non
&&{\dis\hspace{1cm}
+\frac{|H_0|^2-|H_1|^2|\mu|^{-1}}2
\sum_{n=0}^\infty \int_0^\infty
\frac{d^nx}{n!}\left(-\frac{|\lambda\mu|}{2\pi}\right)^n
}\non
&&{\dis\hspace{3.5cm}\left.\times
{\det}_n\left[\frac{\hat V(x_j,x_k)}{\sqrt{x_jx_k}}
e^{-\frac12(\theta(x_j)+\theta(x_k))}\right]\right\}\ketd.}
\ea
%
%\newpage
Both of these series have the form \Eq{TdefFrdet}, so  they can
be summed up and written as   determinants of integral operators
(Fredholm determinants)
\ba{DRcordet}
&&{\dis\hspace{-1cm}
\langle0|\co(0,0)\co(x,t)|0\rangle
=\brad\frac1{[s]^2}\left\{\frac{|H_0|^2+|H_1|^2|\mu|^{-1}}2
\det(I+\gamma \hat U)\right.}\non
&&{\dis\hspace{2cm}\left.
+\frac{|H_0|^2-|H_1|^2|\mu|^{-1}}2
\det(I-\gamma \hat U)\right\}\ketd,}
\ea
where
\be{DRU}
\hat U(x,y)=\frac{\hat V(x,y)}{\sqrt{xy}}
e^{-\frac12(\theta(x)+\theta(y))},
\ee
and
\be{DRgamma}
\gamma=\frac4\pi\sin^2\frac{\pi B}2.
\ee

The determinant representation \Eq{DRcordet} is the main result of the paper,
therefore we summarize here the basic definitions.

The integral operators $I\pm\gamma\hat U$ act on a trial function
$f(x)$ as
\be{DRaction}
[(I\pm\gamma\hat U)f](x)=f(x)\pm\gamma\int_0^\infty\,
\hat U(x,y)f(y)dy.
\ee
The kernel $\hat U(x,y)$ is equal to
\be{DRkernel}
\hat U(x,y)=\frac{\hat D(x,y)}{\sqrt{xy}}
e^{-\frac12(\theta(x)+\theta(y))}e^{\frac12(\Phi_0(x)+\Phi_0(y))},
\ee
where
\be{FDjk2}
\hD(x,y)=\frac1{8\pi i}\oint\frac{dw}{w}
\Bigl(\hY(w,x)+\hY(-w,x)\Bigr)
\Bigl(\hY(w^{-1},y)+\hY(-w^{-1},y)\Bigr),
\ee
and
\be{FDhY}
\hY(w,x)=\frac{e^{\Phi_1(w)}}{q-q^{-1}}
\left(\frac{q^s}{wq^{1/2}+x}-\frac{q^{-s}}{wq^{-1/2}+x}\right).
\ee
The dual fields $\Phi_0(x)$ and $\Phi_1(x)$ were defined in the
section \ref{DF} (see \Eq{Ddef} and \Eq{Ddef0}).
 The main property of these dual fields is,
that they commute with each other, so the Fredholm determinants
$\det(I\pm\gamma\hat U)$ are well defined. Certainly $\det(I\pm\gamma\hat
U)$ are operators in auxiliary Fock space, but they belong to the 
Abelian sub algebra.
 On the other hand, the vacuum
expectation value of these operators is non-trivial. It follows from 
 commutation relations \Eq{Dcom}, \Eq{Dcom0}, that in order to calculate the
vacuum expectation value, one should use the following 
prescription
\be{FDpresc}
\brad\prod_{a=1}^{M_1}e^{\Phi_0(x_a)}
\prod_{b=1}^{M_2}e^{\Phi_1(x_b)}\ketd=
\prod_{a=1}^{M_1}\prod_{b=1}^{M_1}e^{\frac12\eta(x_a,x_b)}
\prod_{a=1}^{M_1}\prod_{b=1}^{M_2}e^{\xi(x_a,x_b)}.
\ee
Here
\be{FDeta0}
\eta(x,y)=2\log\left|\frac{\fmin{\log\frac xy}}{x^2-y^2}\right|,
\ee
and
\be{FDcom}
\xi(x,y)=\log\Bigl((x+yq^{1/2})(x+yq^{-1/2})\Bigr).
\ee

Recall also, that the determinant representation \Eq{DRcordet} is
valid for an arbitrary $s$. If $s=0$, one should understand the r.h.s. as
a limit $s\to 0$, taking into account that  $H_0=0$.

Similar Fredholm determinant  representations were useful not only in the
 theory of correlation functions \cite{BMW}--\cite{LLSS}, but also in
matrix  models \cite{D}--\cite{FO}.
A work on determinant representation for correlation
functions led to the discovery of a determinant formula for a partition
function of the six-vertex model with domain wall boundary conditions
\cite{I}. In the paper \cite{O}  it was shown that this partition function
satisfies Hirota equation. In the paper \cite{U}  it was shown that 
the  determinant formula for the partition function of the six-vertex 
model helps to solve a long-standing mathematical problem---to prove 
the alternating sign matrix conjecture.

%%%%%%%%%%%%%%%%%%%%%%%%%%%%%%%%%%%%%%%%%%%%%%%%%%%%%%%%%%%%%
\section{Large $r$-asymptotic \label{LTDA}}
%%%%%%%%%%%%%%%%%%%%%%%%%%%%%%%%%%%%%%%%%%%%%%%%%%%%%%%%%%%%%

In this section we shall demonstrate, how one can find a long distance
asymptotic of a correlation function starting from the Fredholm
determinant. We shall reproduce some known results.

The kernel of the integral operator $\hat U(x,y)$ can be written in
the form
\be{Lkern}
\hat U(x,y)=\oint\, dwP_1(w,x)P_2(w,y),
\ee
where projectors $P$ are
\be{LP1}
P_1(w,x)=\frac1{8\pi iw\sqrt x}\Bigl(
\hY(w,x)+\hY(-w,x)\Bigr)e^{\frac12\Phi_0(x)-
\frac12\theta(x)},
\ee
\be{LP2}
P_2(w,y)=\frac1{\sqrt y}\Bigl(
\hY(w^{-1},y)+\hY(-w^{-1},y)\Bigr)e^{\frac12\Phi_0(y)-
\frac12\theta(y)}.
\ee
Let us remind here that $w$ integration goes along a large contour around zero
in positive direction. A radius of the contour should be greater then 1.
The Fredholm determinants of the kernels of type \Eq{Lkern} can
be written as determinants of operators acting in the space
of variables ``$w$''
\be{Lequiv}
\det(I\pm\gamma\hat U(x,y))=\det(I\pm\gamma\tilde U(w_1,w_2)),
\ee
where
\be{LtU}
\tilde U(w_1,w_2)=\int_0^\infty\,dxP_1(w_1,x)P_2(w_2,x).
\ee
The integral operator $\tilde U(w_1,w_2)$ acts on a trial function
$f(w)$ as
\be{Lact}
[(I+\tilde U)f](w_1)=f(w_1)+
\oint\tilde U(w_1,w_2)f(w_2)\,dw_2.
\ee

Consider the case $r\to\infty$. Then the
value of the integral in \Eq{LtU} can be estimated by means of a
steepest descent method. The saddle point of  the function $\theta(x)$
is $x=x_0=1$. Examination of commutation relations of dual
fields \Eq{FDpresc} shows that dual fields can be considered as 
analytic functions in the vicinity of real axis.
Hence, we can estimate the integral in \Eq{LtU} as
\be{LtUest}
\tilde U(w_1,w_2)=
P_1(w_1,1)P_2(w_2,1)\left(\sqrt{\frac{2\pi}{mr}}+O(r^{-3/2})\right).
\ee
Thus, for the large $r$
asymptotics the ker\-nel $\tilde U(w_1,w_2)$ be\-comes
a one-di\-mensi\-onal projector, and its Fredholm determinant is 
equal to 
\be{Lasydet} \det(I\pm\gamma\tilde U)
\to1\pm\gamma\oint\,dw \tilde U(w,w).  
\ee 

In order to calculate a vacuum expectation value of $\tilde U(w,w)$
one can use prescription \Eq{FDpresc}, however it is better to write
down the dual field $\Phi_0(x)$ in terms of the original fields
$\Phi_0(x)=\tilde\Phi_0(x)+\Phi_2(x)$. Then the contribution of
the fields $\tilde\Phi_0(1)$ gives 
\be{LtildePhi}
\brad e^{\tilde\Phi_0(1)}\ketd=e^{\frac12\eta(1,1)}=\lambda^{-1}.
\ee
To find a contribution of the fields $\Phi_1(w)$ and $\Phi_2(x)$ we
can use  \Eq{Dvac1} and \Eq{TJ}
\ba{LcontPhi}
&&{\dis\hspace{-1cm}
\brad \tilde U(w,w)\ketd=
\frac{e^{-mr}}{8\pi i\lambda w}\sqrt{\frac{2\pi}{mr}}
\Bigl(Y_1(w,1)+Y_1(-w,1)\Bigr)}\non
&&{\dis\hspace{3cm}
\times
\Bigl(Y_1(w^{-1},1)+Y_1(-w^{-1},1)\Bigr),}
\ea
where
\ba{LY1}
&&{\dis\hspace{-1cm}
Y_1(w,1)=\frac{(wq^{1/2}+1)(wq^{-1/2}+1)}{q-q^{-1}}
\left(\frac{q^s}{wq^{1/2}+1}-\frac{q^{-s}}{wq^{-1/2}+1}\right)}\non
&&{\dis\hspace{5cm}=
[s]+[s-1/2]w.}
\ea
After substituting this into \Eq{Lasydet} it becomes clear that only
a pole at $w=0$ contributes into the integral, so we  
arrive at
\be{LvacU}
\brad\oint\,dw\tilde U(w,w)\ketd=[s]^2\lambda^{-1}
\sqrt{\frac{2\pi}{mr}}e^{-mr},
\ee
and hence
\be{Lvacdet}
\brad\det(I\pm\gamma\tilde U)\ketd\to1\pm[s]^2\frac\gamma\lambda
\sqrt{\frac{2\pi}{mr}} e^{-mr}.
\ee
Finally, substituting this into \Eq{DRcordet}, and using
explicit expressions for $\lambda$ \Eq{DRxi},
$\mu$ \Eq{DRconst} and  $\gamma$ \Eq{DRgamma} we obtain the correct
asymptotical  expression
\be{Lasycor}
\langle0|\co(0,0)\co(x,t)|0\rangle\to\frac{|H_0|^2}{[s^2]}
+|H_1|^2(2\pi mr)^{-1/2}e^{-mr}.
\ee
Recall, that for the correlation
function of local fields one should put $H_0=0=H_0/[s]$,
 therefore we see, that asymptotic formula  \Eq{Lasycor} is
well defined for arbitrary $s$.
If $H_1=0$ (for the stress-energy tensor), then \Eq{Lasycor} gives
 a constant for an asymptotic. However, in this case one had to
estimate the kernel $\tilde U(w_1,w_2)$ more accurately. Namely, one
should take into account explicit expression for corrections of order
$r^{-3/2}$ in \Eq{LtUest}. In this case the kernel $\tilde U$ turns
into two-dimensional projector and it is easy to show, that exponentially
decreasing term behaves like $\exp(-2mr)$.

%%%%%%%%%%%%%%%%%%%%%%%%%%%%%%%%%%%%%%%%%%%%%%
\section*{Summary}
%%%%%%%%%%%%%%%%%%%%%%%%%%%%%%%%%%%%%%%%%%%%%%
 We was able to sum up contributions of all the form factors and to obtain
the closed expression for correlation functions \Eq{DRcordet}.
We believe that it will be possible to do not only for  Toda models
\cite{MOP} but also for all models of integrable  
massive relativistic field theory.

%%%%%%%%%%%%%%%%%%%%%%%%%%%%%%%%%%%%%%%%%%%%%%
\section*{Acknowledgments}
%%%%%%%%%%%%%%%%%%%%%%%%%%%%%%%%%%%%%%%%%%%%%%%
We are grateful to A.~Berkovich, G.~Mussardo, F.~Lessage, S.~Lukyanov
and T.Oota for useful discussions.
This work was supported in part by NSF under Grant No. PHY-9605226,  
by the RFBR under Grant No. 96-01-00344 and INTAS-93-1038.
%%%%%%%%%%%%%%%%%%%%%%%%%%%%%%%%%%%%%%%%%%%%%%%%%%%%%%%%%%%%%%

\newpage

\appendix

%%%%%%%%%%%%%%%%%%%%%%%%%%%%%%%%%%%%%%%%%%%%%%%%%%%%%%%%%%
\section{Elementary symmetric polynomials \label{ESP}}
%%%%%%%%%%%%%%%%%%%%%%%%%%%%%%%%%%%%%%%%%%%%%%%%%%%%%%%%%%
Consider an integral representation \Eq{Tsympol} for the elementary
symmetric polynomials
\be{Esympol}
\sigma_k^{(n)}(x_1,\dots,x_n)\equiv \sigma_k=
\frac1{2\pi i}\oint_{|z|=\rho}\frac {dz}{z^{n-k+1}}
\prod\limits_{m=1}^n(z+x_m),
\ee
where $\rho$ is arbitrary positive. Notice, that the representation
\Eq{Esympol} holds for arbitrary integer $k$ including $k<0$ and $k>n$.

Here we are deriving auxiliary formula, which is used in the section 
\ref{TFF}. Namely,
let us consider a  sum
\be{Esum1}
T_{jk}(\alpha)=\sum_{i=1}^n\alpha^{2i}
\sigma_{2i-j}\sigma_{2i-k},
\qquad j,k=1,\dots,n,
\ee
where  $\alpha$ is an arbitrary complex number.
One can extend the summation in \Eq{Esum1} from $-\infty$ to $n$. Then
we have
\be{Esum2}
T_{jk}(\alpha)=\sum_{i=-\infty}^n
\alpha^{2i}\sigma_{2i-j}\sigma_{2i-k}
=\sum_{l=0}^\infty\alpha^{2(n-l)}\sigma_{2n-2l-j}\sigma_{2n-2l-k}.
\ee
Using the integral representation \Eq{Esympol} we find
\be{Eint}
T_{jk}(\alpha)=\frac1{(2\pi i)^2}\sum_{l=0}^\infty
\oint \frac{d^2z\alpha^{2n-2l}}{z_1^{2l+j-n+1}z_2^{2l+k-n+1}}
\prod\limits_{m=1}^n(z_1+x_m)(z_2+x_m).
\ee
We  can choose the integration contour in such a way
that $|\alpha z_1z_2|>1$ at the contour. Then one can sum up the series with
respect to $l$:
\be{Eint1}
T_{jk}(\alpha)=\frac{\alpha^{2n+2}}{(2\pi i)^2}
\oint\,d^2z \frac{z_1^{n-j+1}z_2^{n-k+1}}{\alpha^2z_1^2z_2^2-1}
\prod\limits_{m=1}^n(z_1+x_m)(z_2+x_m).
\ee
The integrand contains only two simple poles $\alpha z_1z_2=\pm1$,
therefore one can take the integral  with respect to $z_1$ or $z_2$
and  obtain a single integral expression for $T_{jk}$.

%%%%%%%%%%%%%%%%%%%%%%%%%%%%%%%%%%%%%%%%%%%%%%%%%%%%%%%%%%%%%%%%%%
\section{Properties of the Vandermonde matrix \label{SUD}}
%%%%%%%%%%%%%%%%%%%%%%%%%%%%%%%%%%%%%%%%%%%%%%%%%%%%%%%%%%%%%%%%%%

Consider a Vandermonde matrix $W_{jk}$:
\be{SVand}
W_{jk}=z_j^{k-1},\qquad j,k=1,\dots,n,
\ee
with
\be{SdetVand}
{\det}_n (W_{jk})=\prod\limits_{a>b}^n(z_a-z_b).
\ee
The inverse matrix $W^{-1}$ can be written in a form
\be{SinvVand}
(W^{-1})_{jk}=\frac1{(j-1)!}\left.\frac{d^{j-1}}{dx^{j-1}}
\prod\limits_{m\ne k}^n\left(\frac{x-z_m}{z_k-z_m}\right)
\right|_{x=0}.
\ee
Indeed
\be{Ssum}
\sum_{l=1}^nW_{jl}(W^{-1})_{lk}=
\sum_{l=0}^{n-1}\frac{z_j^{l}}{l!}\left.\frac{d^{l}}{dx^{l}}
\prod\limits_{m\ne k}^n\left(\frac{x-z_m}{z_k-z_m}\right)
\right|_{x=0}=\prod\limits_{m\ne k}^n\left(\frac{z_j-z_m}{z_k-z_m}\right)
=\delta_{jk}.
\ee
Here we have used the fact, that the r.h.s. of \Eq{Ssum} is a
Tailor series for the polynomial of the $(n-1)$ degree
$\prod_{m\ne k}^n(x-z_m)(z_k-z_m)^{-1}$.

In the section \ref{TFF} we used the matrix $A_{jk}$:
\be{SA}
A_{jk}=\frac1{(n-j)!}\left.\frac{d^{n-j}}{dx^{n-j}}
\prod\limits_{m\ne k}^n\left(x+x_m\right)
\right|_{x=0}.
\ee
The determinant of this matrix is equal to
\be{SdetA}
{\det}_n A=\prod\limits_{a\ne b}^n(x_a-x_b)
\det\left[A_{jk}\prod\limits_{m\ne k}(x_m-x_k)^{-1}\right].
\ee
It is easy to see that the matrix in the r.h.s. of \Eq{SdetA} coincide
with inverse Vandermonde matrix $W^{-1}$ up to replacement $x_m=-z_m$
and a permutation of rows. Thus, we obtain
\be{SdetA1}
{\det}_n A=\prod\limits_{a<b}^n(x_a-x_b).
\ee

A calculation of  products of the matrix $A$ and matrices, containing
powers of some complex numbers $w$, is  simple. For example,
deriving \Eq{TDjk} we  used
\ba{SUTs}
&&{\dis\hspace{-1cm}
\sum_{l=1}^n\frac1{(n-l)!}\left.\frac{d^{n-l}}{dx^{n-l}}
\prod_{m\ne j}^n(x+x_m)\right|_{x=0}\cdot w_1^{n-l}
G^{(l)}(w_1)}\non
&&{\dis\hspace{-1cm}
=\frac{q^s}{q-q^{-1}}
\prod_{m=1}^n(w_1q^{-1/2}+x_m)
\sum_{l=1}^n\frac{(w_1q^{1/2})^{n-l}}{(n-l)!}
\left.\frac{d^{n-l}}{dx^{n-l}}
\prod_{m\ne j}^n(x+x_m)\right|_{x=0}}\non
&&{\dis\hspace{-1cm}
-\frac{q^{-s}}{q-q^{-1}}
\prod_{m=1}^n(w_1q^{1/2}+x_m)
\sum_{l=1}^n\frac{(w_1q^{-1/2})^{n-l}}{(n-l)!}
\left.\frac{d^{n-l}}{dx^{n-l}}
\prod_{m\ne j}^n(x+x_m)\right|_{x=0}}\non
&&{\dis\hspace{-1cm}
=\frac{
\prod\limits_{m=1}^n(w_1q^{1/2}+x_m)(w_1q^{-1/2}+x_m)}{q-q^{-1}}
\left[\frac{q^s}{w_1q^{1/2}+x_j}-\frac{q^{-s}}{w_1q^{-1/2}+x_j}
\right].}
\ea

\end{document}